\documentclass[twocolumn,amsmath,amssymb,prb,aps]{revtex4}

\usepackage{graphicx}
\usepackage{dcolumn}
\usepackage{bm}
\usepackage{amsmath}
\usepackage{amsfonts}
\usepackage{amssymb}
\usepackage{MnSymbol}

\begin{document}


\title{High-speed phonon imaging using frequency-multiplexed kinetic inductance detectors}

\author{L.~J. Swenson$^{1}$}
\author{A. Cruciani$^{1,2}$}
\author{A. Benoit$^1$}
\author{M. Roesch$^3$}
\author{C.~S. Yung$^4$}
\author{A. Bideaud$^1$}
\author{A. Monfardini$^1$}
\affiliation{$^1$Institut N\'eel, CNRS \& Universit\'{e} Joseph Fourier, BP 166, 38042 Grenoble, France}
\affiliation{$^2$Dipartimento di Fisica, Universit\'a di Roma La Sapienza, p.le A. Moro 2, 00185 Roma, Italy}
\affiliation{$^3$Institut de Radio Astronomie Millim\'etrique, 300 rue de la Piscine, 38406 Saint Martin d'H\`eres, France}
\affiliation{$^4$Superconductor Technologies Inc., 460 Ward Drive, Santa Barbara, CA, 93111, United States}

\date{\today}

\begin{abstract}
    We present a measurement of phonon propagation in a silicon wafer utilizing an array of frequency-multiplexed superconducting resonators coupled to a single transmission line.  The electronic readout permits fully synchronous array sampling with a per-resonator bandwidth of 1.2 MHz, allowing sub-$\mu$s array imaging.  This technological achievement is potentially vital in a variety of low-temperature applications, including single-photon counting, quantum-computing and dark-matter searches.
\end{abstract}


\maketitle

Superconducting microwave resonators are a promising detection technology for a variety of low-temperature applications.  Electromagnetic resonators have been coupled to quantum information circuits\cite{Wallraff2004} and nanomechanical systems\cite{Regal2008}. They are also easily integrated with an absorber or an antenna, enabling them to be used for photon-limited astronomy, x-ray spectroscopy, or in dark-matter searches\cite{mazin:222507, Golwala2008}.  Their tunable resonance frequencies, narrow bandwidths and compatibility with large-bandwidth cryogenic amplifiers also admit the possibility of frequency-domain multiplexing.  This wire-sharing technique is widely recognized as being crucial to realizing large array sizes since the number of wires accessing the coldest stages of a cryostat is necessarily limited\cite{mates:023514}.  Frequency-domain multiplexing of a resonator design known as the Kinetic Inductance Detector (KID)\cite{Day2003} has already been demonstrated at low read-out speeds and used in mm-wavelength astronomy\cite{monfardini:2010, swenson:84, yates:042504}.

Despite their utility, the full potential of superconducting resonators has remained unrealized.  Single-resonator measurements typically use fast commercial electronics which give a large measurement bandwidth but are not compatible with array read out.  On the other hand, customized array readouts utilizing programmable digital logic have not achieved per-resonator bandwidths exceeding $\sim$100 Hz.  Here we present the results from a recent experiment which realizes a large 1.2 MHz per-resonator bandwidth for an array of KIDs.  The array is used to resolve the propagation of phonons created by the interaction of a cosmic-ray with the underlying substrate.  Direct application of this phonon imaging sensor is immediately useful for cosmic-ray detection\cite{cobb:092002}, dark matter studies\cite{Ogburn2006} and for characterizing phonon propagation in exotic materials\cite{Chiu2005}.  Further, by surmounting the technological hurdle of array readout with a large per-resonator bandwidth, this measurement lays the ground work for developing, for example, a scalable quantum-computing architecture or single-photon camera based on superconducting resonators.

\begin{figure}[ht]
   \centering
    \includegraphics[width=82mm]{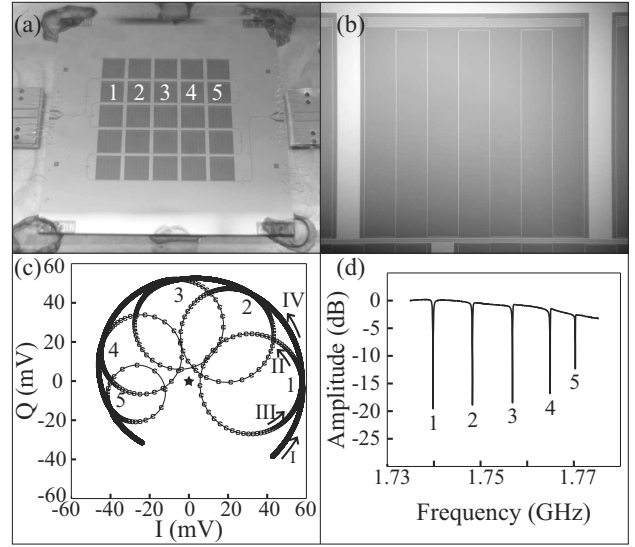}
      \caption{(a) Array micrograph.  Each line of five pixels was designed to occupy a 50 MHz bandwidth. The five labeled pixels were used to measure phonon propagation.  (b) 2 mm x 2 mm LEKID.  (c) In-phase (I) and Quadrature (Q) transmission amplitudes.  The numeric labels correspond to the pixels in (a), the arrows indicate the direction of increasing frequency, and the star ($\star$) marks the origin. (d) Total transmitted amplitude for the measurement shown in (c), calculated using Amplitude$^2$ = I$^2$ + Q$^2$.
      }
         \label{DeviceImages}
\end{figure}

The operating principles of KID read out has been previously discussed\cite{Day2003}.  Briefly, the resonant frequency of an electromagnetic resonator is principally set by the geometric inductance and capacitance.  In a superconductor however, energy can not only be stored in the electromagnetic field, but also in the kinetic energy of the superconducting pairs.  The resulting reactance is known as the kinetic inductance $L_K$.  A superconducting resonator designed to maximize $L_K$ will be sensitive to the internal superconducting state.  Typically, KIDs are designed to absorb photons or substrate phonons with energy exceeding the superconducting gap.  These can then break superconducting pairs and produce a measurable signal.

For the current measurement, a KID geometry known as the the Lumped Element Kinetic Inductance Detector (LEKID) was utilized\cite{doyle:156}. A LEKID consists of a long inductive meander connected to an interdigitated capacitor used to tune the resonant frequency.  An advantage of this geometry is that the current in the inductive section is uniform, resulting in a homogenous response to absorbed phonons throughout the meander.  A micrograph of an array is shown in Fig.\ \ref{DeviceImages}(a).  To fabricate this array, a single 40 nm evaporation of aluminum deposited on 270 $\mu$m thick silicon.  The film was then patterned using standard UV lithography followed by wet etching.  After dicing, the sample was affixed inside a superconducting enclosure and connected to 50 $\Omega$ coplanar-waveguide with wire bonds.  The sample holder was subsequently mounted in a 100 mK dilution cryostat.

The measured In-phase (I) and Quadrature (Q) transmission amplitudes for a 50 MHz frequency sweep of the array is shown in Fig.\ \ref{DeviceImages}(c).  The total transmitted amplitude ($|$S$_{21}|$) is shown in Fig.\ \ref{DeviceImages}(d).  From this plot, the average loaded quality factor Q$_L$ is 1.3$\times 10^4$, limited by the coupling.  By decreasing the coupling, Q$_L$ can be increased up to the internal material-limited quality factor Q$_i$ ($\sim 10^5$).  This affords an increase in sensitivity at the expense of a decreased dynamic range.  For the current measurement, a large dynamic-range was desired in order to differentiate between small and large events.  The resonators were thus somewhat over-coupled.

\begin{figure}[ht]
   \centering
    \includegraphics[width=82mm]{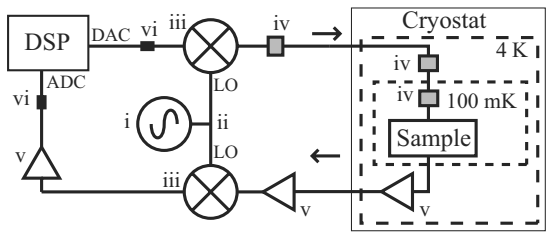}
      \caption{Electronic readout.  A Digital-Signal Processing (DSP) card is used to frequency-multiplex the resonators. The individual components shown are: i) high-frequency synthesizer (.1-8 GHz), ii) splitter, iii) mixer, iv) attenuator, v) amplifier, and vi) low-pass filter.
              }
         \label{Readout}
\end{figure}

Programmable digital electronics were used to perform a fully synchronous transmission measurement of the array.  The details of this technique have previously been discussed\cite{swenson:84}.  For the current measurement, a 12-bit, 100 MSPS ADC was used for sampling the transmitted waveform.  This was followed by digital mixing, low-pass filtering and decimation resulting in a pair of I and Q values for each resonator.  Due to the slow roll-off of the digital 1.2 MHz low-pass filter implemented, an inter-resonator spacing of at least $\sim$4 MHz was necessary to eliminate potential interference.  Along with the 50 MHz Nyquist bandwidth of the ADC, the current electronics were thus able to measure up to 12 resonance in parallel.  Using a faster ADC and improved digital filter, $>$100 pixels per transmission line should be achievable with modern electronics.  In order to reduce the high data-rate resulting from fast sampling, an individual trigger for each resonance was established.  On-board memory simultaneously captured all of the signals if any trigger was exceeded.  A diagram of the measurement setup can be seen in Fig.\ \ref{Readout}.

\begin{figure}[ht]
   \centering
    \includegraphics[width=82mm]{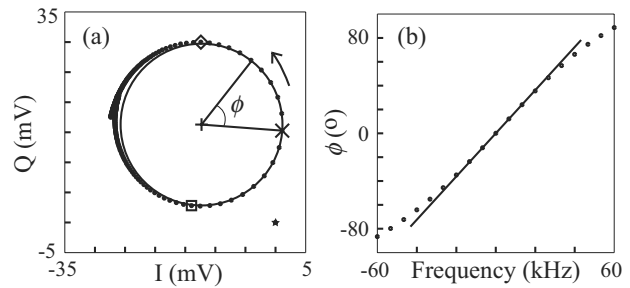}
      \caption{Calibration. (a) The phase $\phi$ as measured around the resonance curve in the IQ plane.  The solid curve is a circular fit to the curvature near the resonance.  The start($\square$) and end ($\diamond$) points of the frequency sweep in (b) are displayed.  (b) Frequency-dependence of $\phi$ for a 120 kHz sweep around the resonance frequency.
              }
         \label{Nika47_CircleFitting}
\end{figure}

The resonant frequency for an unloaded LEKID is given by
  \begin{equation}\label{eq03}
    f_0 = \frac{1}{2 \pi \sqrt{(L_K+L_G) C(\epsilon)}},
 \end{equation}
where $\epsilon$ is the effective surrounding permittivity, and $L_K$, $L_G$ and $C$ are the resonator's kinetic inductance, geometric inductance and capacitance respectively\cite{doyle:156}.  A change in the resonant frequency $\delta f_0$ may be expressed as $\delta f_0 = \delta L_K \partial f_0 / \partial L_K$. From this, $\delta f_0 = (-C f_0^3/2) \delta L_K$.  A small linear shift in the kinetic inductance results in a proportional frequency shift of the resonance feature.  In the IQ plane, a small shift in frequency primarily results in rotation around the resonance curve.  This can be expressed as a rotation angle $\phi$ as shown in Fig.\ \ref{Nika47_CircleFitting}(a).

Since all the resonators are geometrically identical, except for slight differences in the capacitive section introduced to adjust the resonance frequencies, they all posses the same $L_K$.  Hence, for a fixed energy input, all the resonators will experience the same frequency shift $\Delta f$.  In Fig.\ \ref{Nika47_CircleFitting}(b), a plot of $\phi$ is shown for a 120 kHz frequency sweep around a resonance.  A fit to the central data in this plot yields the slope $\Delta \phi / \Delta f$. For small $\phi$, dividing by $\Delta \phi / \Delta f$ results in a corresponding $\Delta f$ and is effectively a calibration of the resonance.

\begin{figure}[ht]
   \centering
    \includegraphics[width=82mm]{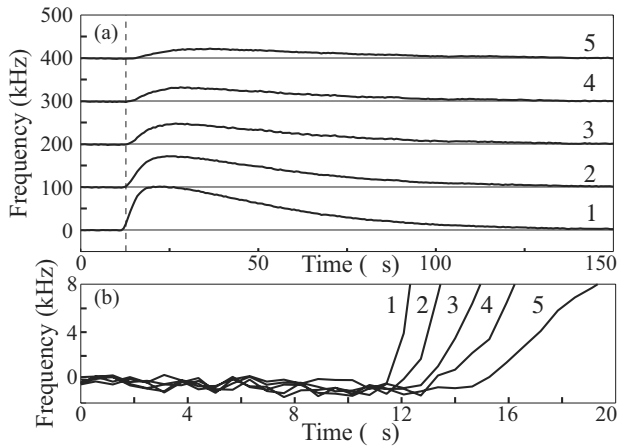}
      \caption{(a) Calibrated high-speed detection of a cosmic event using 5 KIDs. The individual responses are offset by 100 kHz.  The dashed vertical line is the time that a resonator trigger level was exceeded, in this case by pixel 1. (b) Zoom of initial rise clearly showing the propagation delay.
              }
         \label{Event01}
\end{figure}

During a typical measurement, hundreds of events were captured over the course of a few hours at a rate of one event every $\sim$5-20 seconds.  An event causing a moderate response from all 5 pixels is shown in Fig.\ \ref{Event01}.  The decreasing amplitudes in going from pixel 1 to 5 along with the increasing time delay corroborates the hypothesis that the cosmic rays are not interacting directly with the resonators but rather create phonons in the substrate which propagate to the LEKIDs.  At T=100 mK $\ll$ T$_c\sim$1.2 K, thermal phonons do not have sufficient energy to break superconducting pairs and cannot result in a measurable signal.  The phonons which are detected are thus non-thermalized and posses energies exceeding the gap energy.

Collision-induced phonon propagation in silicon has been extensively studied.\cite{ogburn:2008, mandic:509}  In the simplest model, nuclear or electronic recoil from a particle interaction creates a cloud of high energy phonons ($>$5 THz).  Rapid inelastic scattering results in anharmonic down-conversion until reaching a bottleneck frequency of $\sim$1.6 THz at which point they propagate outward, undergoing frequent elastic scattering.  This quasi-diffusive processes is isotropic due to rapid conversion between phonon modes.  The time difference of arrival between pixels and the frequency shifts thus enables a determination of the initial event position and the propagation velocity.  For the event in Fig.\ \ref{Event01}, the position was determined to be at (-6.3$\pm$.3 mm,$\pm$3.6$\pm$.3 mm), where (0,0) is located at the geometric center of the array and the x-axis lies along the pixel centers.  The degeneracy in the y-coordinate is due to the symmetry of the array which can be easily broken in future measurements.  The positional error is due to the arrival time uncertainty, equal to the inverse per-resonator bandwidth, and the $\sim$2 kHz noise in the measured frequency shift.  Along with the thin substrate and complicating surface effects, the positional error on the order of the wafer thickness obscures any potential depth information for the current measurement.  The propagation velocity was determined to be 2.2$\pm$.9 km/s.

By fitting the low frequency-shift tails of the cosmic events to an exponential, a time constant of 36 $\mu$s was measured.  This significantly exceeds the resonator time constant $\tau =($Q$_L)$/$f_0$ = 7 $\mu$s, which is likely due to the long quasi-particle decay time in aluminum at T$\ll$T$_c$.  Finally, the initial rise of the event is caused by a parametric shift in the resonant circuit which is not limited by $\tau$.  The required resolution of the leading edge thus determines the minimum per-pixel bandwidth for a measurement.  For example, in the current measurement, in order to resolve a $\sim \mu$s time delay between pixels, a per-pixel bandwidth exceeding 1 MHz was critical.

In conclusion, we have imaged phonon propagation with sub-$\mu$s time resolution using frequency-multiplexed superconducting resonators.  A measurement is planned in the near future utilizing a calibrated particle source in order to asses the detector sensitivity.  An improved digital low-pass filter design, a larger bandwidth ADC, smaller pixels, and an increased pixel count will also be implemented in order to decrease the measurement uncertainty.  Based upon these results, a full-scale detector could then be developed for dark matter searches or cosmic-ray detection.  This technique should also prove invaluable in developing a scalable quantum-computing architecture or single-photon camera based on superconducting resonators.

\begin{acknowledgments}
      We acknowledge Simon Doyle, Karl Schuster, Christian Hoffmann, Julien Minet and Christopher Moon for useful discussions.  Part of this work was supported by grant ANR-09-JCJC-0021-01 of the French National Research Agency, the Nanosciences Foundation of Grenoble and R\'egion Rh\^one-Alpes (program CIBLE 2009).
\end{acknowledgments}


\begin{thebibliography}{15}
\expandafter\ifx\csname natexlab\endcsname\relax\def\natexlab#1{#1}\fi
\expandafter\ifx\csname bibnamefont\endcsname\relax
  \def\bibnamefont#1{#1}\fi
\expandafter\ifx\csname bibfnamefont\endcsname\relax
  \def\bibfnamefont#1{#1}\fi
\expandafter\ifx\csname citenamefont\endcsname\relax
  \def\citenamefont#1{#1}\fi
\expandafter\ifx\csname url\endcsname\relax
  \def\url#1{\texttt{#1}}\fi
\expandafter\ifx\csname urlprefix\endcsname\relax\def\urlprefix{URL }\fi
\providecommand{\bibinfo}[2]{#2}
\providecommand{\eprint}[2][]{\url{#2}}

\bibitem[{\citenamefont{Wallraff et~al.}(2004)\citenamefont{Wallraff, Schuster,
  Blais, Frunzio, Huang, Majer, Kumar, Girvin, and Schoelkopf}}]{Wallraff2004}
\bibinfo{author}{\bibfnamefont{A.}~\bibnamefont{Wallraff}},
  \bibinfo{author}{\bibfnamefont{D.~I.} \bibnamefont{Schuster}},
  \bibinfo{author}{\bibfnamefont{A.}~\bibnamefont{Blais}},
  \bibinfo{author}{\bibfnamefont{L.}~\bibnamefont{Frunzio}},
  \bibinfo{author}{\bibfnamefont{R.-S.} \bibnamefont{Huang}},
  \bibinfo{author}{\bibfnamefont{J.}~\bibnamefont{Majer}},
  \bibinfo{author}{\bibfnamefont{S.}~\bibnamefont{Kumar}},
  \bibinfo{author}{\bibfnamefont{S.~M.} \bibnamefont{Girvin}},
  \bibnamefont{and} \bibinfo{author}{\bibfnamefont{R.~J.}
  \bibnamefont{Schoelkopf}}, \bibinfo{journal}{Nature}
  \textbf{\bibinfo{volume}{431}}, \bibinfo{pages}{162} (\bibinfo{year}{2004}).

\bibitem[{\citenamefont{Regal et~al.}(2008)\citenamefont{Regal, Teufel, and
  Lehnert}}]{Regal2008}
\bibinfo{author}{\bibfnamefont{C.~A.} \bibnamefont{Regal}},
  \bibinfo{author}{\bibfnamefont{J.~D.} \bibnamefont{Teufel}},
  \bibnamefont{and} \bibinfo{author}{\bibfnamefont{K.~W.}
  \bibnamefont{Lehnert}}, \bibinfo{journal}{Nat Phys}
  \textbf{\bibinfo{volume}{4}}, \bibinfo{pages}{555} (\bibinfo{year}{2008}).

\bibitem[{\citenamefont{Mazin et~al.}(2006)\citenamefont{Mazin, Bumble, Day,
  Eckart, Golwala, Zmuidzinas, and Harrison}}]{mazin:222507}
\bibinfo{author}{\bibfnamefont{B.~A.} \bibnamefont{Mazin}},
  \bibinfo{author}{\bibfnamefont{B.}~\bibnamefont{Bumble}},
  \bibinfo{author}{\bibfnamefont{P.~K.} \bibnamefont{Day}},
  \bibinfo{author}{\bibfnamefont{M.~E.} \bibnamefont{Eckart}},
  \bibinfo{author}{\bibfnamefont{S.}~\bibnamefont{Golwala}},
  \bibinfo{author}{\bibfnamefont{J.}~\bibnamefont{Zmuidzinas}},
  \bibnamefont{and} \bibinfo{author}{\bibfnamefont{F.~A.}
  \bibnamefont{Harrison}}, \bibinfo{journal}{Applied Physics Letters}
  \textbf{\bibinfo{volume}{89}}, \bibinfo{pages}{222507}
  (\bibinfo{year}{2006}).

\bibitem[{\citenamefont{Golwala et~al.}(2008)\citenamefont{Golwala, Gao, Moore,
  Mazin, Eckart, Bumble, Day, LeDuc, and Zmuidzinas}}]{Golwala2008}
\bibinfo{author}{\bibfnamefont{S.}~\bibnamefont{Golwala}},
  \bibinfo{author}{\bibfnamefont{J.}~\bibnamefont{Gao}},
  \bibinfo{author}{\bibfnamefont{D.}~\bibnamefont{Moore}},
  \bibinfo{author}{\bibfnamefont{B.}~\bibnamefont{Mazin}},
  \bibinfo{author}{\bibfnamefont{M.}~\bibnamefont{Eckart}},
  \bibinfo{author}{\bibfnamefont{B.}~\bibnamefont{Bumble}},
  \bibinfo{author}{\bibfnamefont{P.}~\bibnamefont{Day}},
  \bibinfo{author}{\bibfnamefont{H.}~\bibnamefont{LeDuc}}, \bibnamefont{and}
  \bibinfo{author}{\bibfnamefont{J.}~\bibnamefont{Zmuidzinas}},
  \bibinfo{journal}{Journal of Low Temperature Physics}
  \textbf{\bibinfo{volume}{151}}, \bibinfo{pages}{550} (\bibinfo{year}{2008}).

\bibitem[{\citenamefont{Mates et~al.}(2008)\citenamefont{Mates, Hilton, Irwin,
  Vale, and Lehnert}}]{mates:023514}
\bibinfo{author}{\bibfnamefont{J.~A.~B.}~\bibnamefont{Mates}},
  \bibinfo{author}{\bibfnamefont{G.~C.}~\bibnamefont{Hilton}},
  \bibinfo{author}{\bibfnamefont{K.~D.}~\bibnamefont{Irwin}},
  \bibinfo{author}{\bibfnamefont{L.~R.}~\bibnamefont{Vale}}, \bibnamefont{and}
  \bibinfo{author}{\bibfnamefont{K.~W.}~\bibnamefont{Lehnert}},
  \bibinfo{journal}{Applied Physics Letters} \textbf{\bibinfo{volume}{92}},
  \bibinfo{pages}{023514} (\bibinfo{year}{2008}).

\bibitem[{\citenamefont{Day et~al.}(2003)\citenamefont{Day, LeDuc, Mazin,
  Vayonakis, and Zmuidzinas}}]{Day2003}
\bibinfo{author}{\bibfnamefont{P.~K.}~\bibnamefont{Day}},
  \bibinfo{author}{\bibfnamefont{H.~G.}~\bibnamefont{LeDuc}},
  \bibinfo{author}{\bibfnamefont{B.~A.}~\bibnamefont{Mazin}},
  \bibinfo{author}{\bibfnamefont{A.}~\bibnamefont{Vayonakis}},
  \bibnamefont{and}
  \bibinfo{author}{\bibfnamefont{J.}~\bibnamefont{Zmuidzinas}},
  \bibinfo{journal}{Nature} \textbf{\bibinfo{volume}{425}},
  \bibinfo{pages}{817} (\bibinfo{year}{2003}).

\bibitem[{\citenamefont{Monfardini et~al.}()\citenamefont{Monfardini, Swenson,
  Bideaud, D\'esert, Yates, Benoit, Baryshev, Baselmans, Doyle, Klein
  et~al.}}]{monfardini:2010}
\bibinfo{author}{\bibfnamefont{A.}~\bibnamefont{Monfardini}},
  \bibinfo{author}{\bibfnamefont{L.~J.}~\bibnamefont{Swenson}},
  \bibinfo{author}{\bibfnamefont{A.}~\bibnamefont{Bideaud}},
  \bibinfo{author}{\bibfnamefont{F.~X.}~\bibnamefont{D\'esert}},
  \bibinfo{author}{\bibfnamefont{S.~J.~C.}~\bibnamefont{Yates}},
  \bibinfo{author}{\bibfnamefont{A.}~\bibnamefont{Benoit}},
  \bibinfo{author}{\bibfnamefont{A.~M.}~\bibnamefont{Baryshev}},
  \bibinfo{author}{\bibfnamefont{J.~J.~A.}~\bibnamefont{Baselmans}},
  \bibinfo{author}{\bibfnamefont{S.}~\bibnamefont{Doyle}},
  \bibinfo{author}{\bibfnamefont{B.}~\bibnamefont{Klein}},
  \bibinfo{author}{\bibfnamefont{M.}~\bibnamefont{Roesch}},
  \bibinfo{author}{\bibfnamefont{C.}~\bibnamefont{Tucker}},
  \bibinfo{author}{\bibfnamefont{P.}~\bibnamefont{Ade}},
  \bibinfo{author}{\bibfnamefont{M.}~\bibnamefont{Calvo}},
  \bibinfo{author}{\bibfnamefont{P.}~\bibnamefont{Camus}},
  \bibinfo{author}{\bibfnamefont{C.}~\bibnamefont{Giordano}},
  \bibinfo{author}{\bibfnamefont{R.}~\bibnamefont{Guesten}},
  \bibinfo{author}{\bibfnamefont{C.}~\bibnamefont{Hoffmann}},
  \bibinfo{author}{\bibfnamefont{S.}~\bibnamefont{Leclercq}},
  \bibinfo{author}{\bibfnamefont{P.}~\bibnamefont{Mauskopf}},
  \bibnamefont{and}
  \bibinfo{author}{\bibfnamefont{K.~F.}~\bibnamefont{Schuster}},
  \bibinfo{note}{arxiv:1004.2209 (unpublished)}.

\bibitem[{\citenamefont{Swenson et~al.}(2009)\citenamefont{Swenson, Minet,
  Grabovskij, Buisson, Lecocq, Hoffmann, Camus, Vill\'{e}gier, Doyle, Mauskopf
  et~al.}}]{swenson:84}
\bibinfo{author}{\bibfnamefont{L.~J.}~\bibnamefont{Swenson}},
  \bibinfo{author}{\bibfnamefont{J.}~\bibnamefont{Minet}},
  \bibinfo{author}{\bibfnamefont{G.~J.}~\bibnamefont{Grabovskij}},
  \bibinfo{author}{\bibfnamefont{O.}~\bibnamefont{Buisson}},
  \bibinfo{author}{\bibfnamefont{F.}~\bibnamefont{Lecocq}},
  \bibinfo{author}{\bibfnamefont{C.}~\bibnamefont{Hoffmann}},
  \bibinfo{author}{\bibfnamefont{P.}~\bibnamefont{Camus}},
  \bibinfo{author}{\bibfnamefont{J.-C.}~\bibnamefont{Vill\'{e}gier}},
  \bibinfo{author}{\bibfnamefont{S.}~\bibnamefont{Doyle}},
  \bibinfo{author}{\bibfnamefont{P.}~\bibnamefont{Mauskopf}},
  \bibinfo{author}{\bibfnamefont{M.}~\bibnamefont{Roesch}},
  \bibinfo{author}{\bibfnamefont{M.}~\bibnamefont{Calvo}},
  \bibinfo{author}{\bibfnamefont{C.}~\bibnamefont{Giordano}},
  \bibinfo{author}{\bibfnamefont{S.~J.~C.}~\bibnamefont{Yates}},
  \bibinfo{author}{\bibfnamefont{A.~M.}~\bibnamefont{Baryshev}},
  \bibinfo{author}{\bibfnamefont{J.~J.~A.}~\bibnamefont{Baselmans}},
  \bibinfo{author}{\bibfnamefont{A.}~\bibnamefont{Benoit}},
  \bibnamefont{and}
  \bibinfo{author}{\bibfnamefont{A.}~\bibnamefont{Monfardini}},
  \bibinfo{journal}{in The Thirteenth International
  Workshop On Low Temperature Detectors (LTD-13), AIP Proc.}
  \textbf{\bibinfo{volume}{1185}}, \bibinfo{pages}{84} (\bibinfo{year}{2009}).

\bibitem[{\citenamefont{Yates et~al.}(2009)\citenamefont{Yates, Baryshev,
  Baselmans, Klein, and G\"{u}sten}}]{yates:042504}
\bibinfo{author}{\bibfnamefont{S.~J.~C.} \bibnamefont{Yates}},
  \bibinfo{author}{\bibfnamefont{A.~M.} \bibnamefont{Baryshev}},
  \bibinfo{author}{\bibfnamefont{J.~J.~A.} \bibnamefont{Baselmans}},
  \bibinfo{author}{\bibfnamefont{B.}~\bibnamefont{Klein}}, \bibnamefont{and}
  \bibinfo{author}{\bibfnamefont{R.}~\bibnamefont{G\"{u}sten}},
  \bibinfo{journal}{Applied Physics Letters} \textbf{\bibinfo{volume}{95}},
  \bibinfo{pages}{042504} (\bibinfo{year}{2009}).

\bibitem[{\citenamefont{Cobb et~al.}(2000)\citenamefont{Cobb, Marshak, Allison,
  Alner, Ayres, Barrett, Bode, Border, Brooks, Cotton et~al.}}]{cobb:092002}
\bibinfo{author}{\bibfnamefont{J.~H.} \bibnamefont{Cobb}},
  \bibinfo{author}{\bibfnamefont{M.~L.} \bibnamefont{Marshak}},
  \bibinfo{author}{\bibfnamefont{W.~W.~M.} \bibnamefont{Allison}},
  \bibinfo{author}{\bibfnamefont{G.~J.} \bibnamefont{Alner}},
  \bibinfo{author}{\bibfnamefont{D.~S.} \bibnamefont{Ayres}},
  \bibinfo{author}{\bibfnamefont{W.~L.} \bibnamefont{Barrett}},
  \bibinfo{author}{\bibfnamefont{C.}~\bibnamefont{Bode}},
  \bibinfo{author}{\bibfnamefont{P.~M.} \bibnamefont{Border}},
  \bibinfo{author}{\bibfnamefont{C.~B.} \bibnamefont{Brooks}},
  \bibinfo{author}{\bibfnamefont{R.~J.} \bibnamefont{Cotton}},
  \bibinfo{author}{\bibfnamefont{H.}~\bibnamefont{Courant}},
  \bibinfo{author}{\bibfnamefont{D.~M.}~\bibnamefont{Demuth}},
  \bibinfo{author}{\bibfnamefont{T.~H.}~\bibnamefont{Fields}},
  \bibinfo{author}{\bibfnamefont{H.~R.}~\bibnamefont{Gallagher}},
  \bibinfo{author}{\bibfnamefont{M.~C.}~\bibnamefont{Goodman}},
  \bibinfo{author}{\bibfnamefont{R.}~\bibnamefont{Gran}},
  \bibinfo{author}{\bibfnamefont{T.}~\bibnamefont{Joffe-Minor}},
  \bibinfo{author}{\bibfnamefont{T.}~\bibnamefont{Kafka}},
  \bibinfo{author}{\bibfnamefont{S.~M.~S.}~\bibnamefont{Kasahara}},
  \bibinfo{author}{\bibfnamefont{W.}~\bibnamefont{Leeson}},
  \bibinfo{author}{\bibfnamefont{P.~J.}~\bibnamefont{Litchfield}},
  \bibinfo{author}{\bibfnamefont{N.~P.}~\bibnamefont{Longley}},
  \bibinfo{author}{\bibfnamefont{W.~A.}~\bibnamefont{Mann}},
  \bibinfo{author}{\bibfnamefont{R.~H.}~\bibnamefont{Milburn}},
  \bibinfo{author}{\bibfnamefont{W.~H.}~\bibnamefont{Miller}},
  \bibinfo{author}{\bibfnamefont{C.}~\bibnamefont{Moon}},
  \bibinfo{author}{\bibfnamefont{L.}~\bibnamefont{Mualem}},
  \bibinfo{author}{\bibfnamefont{A.}~\bibnamefont{Napier}},
  \bibinfo{author}{\bibfnamefont{W.~P.}~\bibnamefont{Oliver}},
  \bibinfo{author}{\bibfnamefont{G.~F.}~\bibnamefont{Pearce}},
  \bibinfo{author}{\bibfnamefont{E.~A.}~\bibnamefont{Peterson}},
  \bibinfo{author}{\bibfnamefont{D.~A.}~\bibnamefont{Petyt}},
  \bibinfo{author}{\bibfnamefont{L.~E.}~\bibnamefont{Price}},
  \bibinfo{author}{\bibfnamefont{K.}~\bibnamefont{Ruddick}},
  \bibinfo{author}{\bibfnamefont{M.}~\bibnamefont{Sanchez}},
  \bibinfo{author}{\bibfnamefont{P.}~\bibnamefont{Sankey}},
  \bibinfo{author}{\bibfnamefont{J.}~\bibnamefont{Schneps}},
  \bibinfo{author}{\bibfnamefont{M.~H.}~\bibnamefont{Schub}},
  \bibinfo{author}{\bibfnamefont{R.}~\bibnamefont{Seidlein}},
  \bibinfo{author}{\bibfnamefont{A.}~\bibnamefont{Stassinakis}},
  \bibinfo{author}{\bibfnamefont{J.~L.}~\bibnamefont{Thron}},
  \bibinfo{author}{\bibfnamefont{V.}~\bibnamefont{Vassiliev}},
  \bibinfo{author}{\bibfnamefont{G.}~\bibnamefont{Villaume}},
  \bibinfo{author}{\bibfnamefont{S.~P.}~\bibnamefont{Wakely}},
  \bibinfo{author}{\bibfnamefont{N.}~\bibnamefont{West}},
  \bibnamefont{and}
  \bibinfo{author}{\bibfnamefont{D.}~\bibnamefont{Wall}},
  \bibinfo{journal}{Phys. Rev. D}
  \textbf{\bibinfo{volume}{61}}, \bibinfo{pages}{092002}
  (\bibinfo{year}{2000}).

\bibitem[{\citenamefont{Ogburn}(2006)}]{Ogburn2006}
\bibinfo{author}{\bibfnamefont{R.~W.} \bibnamefont{Ogburn}},
  \bibinfo{journal}{in Proc. of the Intl. Symposium On Detector Development For
  Particle, Astroparticle And Synchrotron Radiation Experiments (SNIC)} p.
  \bibinfo{pages}{150} (\bibinfo{year}{2006}).

\bibitem[{\citenamefont{Chiu et~al.}(2005)\citenamefont{Chiu, Deshpande,
  Postma, Lau, Mikó, Forró, and Bockrath}}]{Chiu2005}
\bibinfo{author}{\bibfnamefont{H.-Y.} \bibnamefont{Chiu}},
  \bibinfo{author}{\bibfnamefont{V.~V.} \bibnamefont{Deshpande}},
  \bibinfo{author}{\bibfnamefont{H.~W.~C.} \bibnamefont{Postma}},
  \bibinfo{author}{\bibfnamefont{C.~N.} \bibnamefont{Lau}},
  \bibinfo{author}{\bibfnamefont{C.}~\bibnamefont{Mikó}},
  \bibinfo{author}{\bibfnamefont{L.}~\bibnamefont{Forró}}, \bibnamefont{and}
  \bibinfo{author}{\bibfnamefont{M.}~\bibnamefont{Bockrath}},
  \bibinfo{journal}{Phys. Rev. Lett.} \textbf{\bibinfo{volume}{95}},
  \bibinfo{pages}{226101} (\bibinfo{year}{2005}).

\bibitem[{\citenamefont{Doyle et~al.}(2009)\citenamefont{Doyle, Mauskopf,
  Zhang, Withington, Goldie, Glowacka, Monfardini, Swenson, and
  Roesch}}]{doyle:156}
\bibinfo{author}{\bibfnamefont{S.}~\bibnamefont{Doyle}},
  \bibinfo{author}{\bibfnamefont{P.}~\bibnamefont{Mauskopf}},
  \bibinfo{author}{\bibfnamefont{J.}~\bibnamefont{Zhang}},
  \bibinfo{author}{\bibfnamefont{S.}~\bibnamefont{Withington}},
  \bibinfo{author}{\bibfnamefont{D.}~\bibnamefont{Goldie}},
  \bibinfo{author}{\bibfnamefont{D.}~\bibnamefont{Glowacka}},
  \bibinfo{author}{\bibfnamefont{A.}~\bibnamefont{Monfardini}},
  \bibinfo{author}{\bibfnamefont{L.~J.} \bibnamefont{Swenson}},
  \bibnamefont{and} \bibinfo{author}{\bibfnamefont{M.}~\bibnamefont{Roesch}},
  \bibinfo{journal}{in The Thirteenth International Workshop On Low Temperature
  Detectors (LTD-13), AIP Proc.} \textbf{\bibinfo{volume}{1185}},
  \bibinfo{pages}{156} (\bibinfo{year}{2009}).

\bibitem[{\citenamefont{Ogburn}(2008)}]{ogburn:2008}
\bibinfo{author}{\bibfnamefont{R.~W.} \bibnamefont{Ogburn}}, Ph.D. thesis,
  \bibinfo{school}{Stanford University} (\bibinfo{year}{2008}).

\bibitem[{\citenamefont{Mandic et~al.}(2002)\citenamefont{Mandic, Rau, Akerib,
  Brink, Cabrera, Castle, Chang, Crisler, Driscoll, Emes et~al.}}]{mandic:509}
\bibinfo{author}{\bibfnamefont{V.}~\bibnamefont{Mandic}},
  \bibinfo{author}{\bibfnamefont{W.}~\bibnamefont{Rau}},
  \bibinfo{author}{\bibfnamefont{D.}~\bibnamefont{Akerib}},
  \bibinfo{author}{\bibfnamefont{P.}~\bibnamefont{Brink}},
  \bibinfo{author}{\bibfnamefont{B.}~\bibnamefont{Cabrera}},
  \bibinfo{author}{\bibfnamefont{J.~P.} \bibnamefont{Castle}},
  \bibinfo{author}{\bibfnamefont{C.}~\bibnamefont{Chang}},
  \bibinfo{author}{\bibfnamefont{M.~B.} \bibnamefont{Crisler}},
  \bibinfo{author}{\bibfnamefont{D.}~\bibnamefont{Driscoll}},
  \bibinfo{author}{\bibfnamefont{J.}~\bibnamefont{Emes}},
  \bibinfo{author}{\bibfnamefont{R.~J.}~\bibnamefont{Gaitskell}},
  \bibinfo{author}{\bibfnamefont{J.}~\bibnamefont{Hellmig}},
  \bibinfo{author}{\bibfnamefont{M.~E.}~\bibnamefont{Huber}},
  \bibinfo{author}{\bibfnamefont{S.}~\bibnamefont{Kamat}},
  \bibinfo{author}{\bibfnamefont{J.~M.}~\bibnamefont{Martinis}},
  \bibinfo{author}{\bibfnamefont{P.}~\bibnamefont{Meunier}},
  \bibinfo{author}{\bibfnamefont{T.~A.}~\bibnamefont{Perera}},
  \bibinfo{author}{\bibfnamefont{M.}~\bibnamefont{Perillo-Issac}},
  \bibinfo{author}{\bibfnamefont{T.}~\bibnamefont{Saab}},
  \bibinfo{author}{\bibfnamefont{B.}~\bibnamefont{Sadoulet}},
  \bibinfo{author}{\bibfnamefont{R.}~\bibnamefont{Schnee}},
  \bibinfo{author}{\bibfnamefont{D.}~\bibnamefont{Seitz}},
  \bibinfo{author}{\bibfnamefont{G.}~\bibnamefont{Wang}},
  \bibnamefont{and}
  \bibinfo{author}{\bibfnamefont{B.}~\bibnamefont{Young}},
  \bibinfo{journal}{in The Ninth International Workshop On Low Temperature
  Detectors (LTD-9), AIP Proc.} \textbf{\bibinfo{volume}{605}},
  \bibinfo{pages}{509} (\bibinfo{year}{2002}).

\end{thebibliography}

\end{document}